\documentclass[aps,pra,twocolumn,groupedaddress,showpacs]{revtex4}

\usepackage{graphicx}
\usepackage{epstopdf}

\begin{document}

\DeclareGraphicsExtensions{.eps,.EPS,.jpg,.bmp}


\title{State labelling Wannier-Stark atomic interferometers}


\author{B. Pelle, A. Hilico, G. Tackmann, Q. Beaufils, F. Pereira dos Santos}

\email[]{franck.pereira@obspm.fr}
\affiliation{LNE-SYRTE, UMR 8630 CNRS, Observatoire de Paris,
UPMC, 61 avenue de l'Observatoire, 75014 Paris, FRANCE}


\date{\today}

\begin{abstract}
Using cold $^{87}$Rb atoms trapped in a 1D-optical lattice, atomic interferometers involving coherent superpositions between different Wannier-Stark atomic states are realized. Two different kinds of trapped interferometer schemes are presented: a Ramsey-type interferometer sensitive both to clock frequency and external forces, and a symmetric accordion-type interferometer, sensitive to external forces only. We evaluate the limits in terms of sensitivity and accuracy of those schemes and discuss their application as force sensors. As a first step, we apply these interferometers to the measurement of the Bloch frequency and the demonstration of a compact gravimeter.
\end{abstract}


\maketitle



Laser pulse driven atom interferometers (AI) have demonstrated their ability to realize absolute measurements with good accuracy and sensitivity. A key feature in these experiments is the precise control of the separation between the arms, which for free falling atoms arises from the momentum exchange with the laser field. This leads to separations increasing with interrogation time, and accurate knowledge of the scale factor. Manipulations of atomic wavepackets can be realized using various types of beamsplitters, some allowing only for external state changes, such as Bragg diffraction, others being accompanied by changes in the internal state as well, such as Raman transitions. More details on the different beamsplitter tools and interferometer configurations can be found in~\cite{AcadPress_Berman, RevModPhys_Pritchard}.
Among the different types of instruments that have been developed so far, AI based on Raman transitions have been used to realize high precision inertial sensors with state-of-the-art performances, such as gyrometers~\cite{ClassQuantumGrav_Kasevich_Gyro, PRA_Landragin, PRL_Kasevich_Gyro, NJP_Rasel}, gravimeters~\cite{Metrologia_Chu, APB_LeGouet, GyroAndNav_Schmidt} and gradiometers~\cite{PRA_Kasevich_Gradio, PRL_Tino_Gradio}, as well as velocity sensors for the measurement of the photon recoil and determination of the fine structure constant~\cite{PRL_Bouchendira}. Yet, such laser transitions also efficiently operate on trapped atoms, as demonstrated in~\cite{PRL_Beaufils}. They allow for transporting the atoms in a coherent way in a vertical optical lattice and manipulating them in different Wannier-Stark states, that constitute, with a good approximation, the eigenstates of this system. In particular, the realization of a state separated interferometer allows for the realization of a force sensor.
Various schemes have been proposed~\cite{PRA_Hansch, PRD_Dimopoulos, PRL_Inguscio, PRA_Wolf_Fromopticallatticeclock} and demonstrated~\cite{PRL_Tino_SrLattice1, PRL_Tino_SrLattice2, PRL_Hinds, PRL_Tino_SrLattice3, PRL_Tino_SrLattice4, PRA_Bresson} to realize high sensitivity force sensors with confined atoms, based on the manipulation of atomic motional states, using either lasers or magnetic forces~\cite{PRL_Leanhardt, PRL_Wu, Nature_Kruger, PRL_Fortagh, PRA_Sackett, NJP_Meschede, NJP_Anderson}. The trapped geometry allows for long interrogation times~\cite{PRL_Tino_SrLattice4, PRA_Prentiss} and compactness, and offers interesting perspectives as sensors for atom-surface interactions~\cite{PRA_Cornell, PRL_Slama, PRL_Tino_SrLattice4, NJP_Cornish}.
In this paper, we demonstrate and study the performances of atom interferometers based on superpositions of Wannier-Stark states. We first discuss the simple case of a Ramsey-type interferometer. We study the limits in its sensitivity, and some systematic effects related to the trapping lasers. 
We finally demonstrate a symmetric interferometer, as recently proposed in~\cite{PRA_Sophie}, which offers the advantage of being insensitive to clock related frequency shifts.

\section{Description of the system}

Our system is composed of $^{87}$Rb atoms trapped in a vertical 1D-optical lattice far detuned from the atomic resonance. Laser-cooled atoms which are loaded into the lattice are approximated by a two-level internal atomic structure with stable states, defined by the ground $\left|g\right\rangle$ and excited $\left|e\right\rangle$ hyperfine levels separated in energy by $h\nu_{HFS}$. The stationary Hamiltonian of this system in the absence of an external coupling field between internal and external states is given by~\cite{PRA_Thommen, PRA_Wolf_Shallowtrap}:
\begin{equation}
\hat H=\hat H_{int}+\hat H_{K}+\hat H_{l}+\hat H_{g},
\end{equation}
where ${\hat H_{int}=h\nu_{HFS}|e\rangle\langle e|}$ represents the atomic internal energy, ${\hat H_{K}=\frac{\hat P^2}{2 m_a}}$ is the kinetic energy, ${\hat H_{l}=\frac{U_0}{2}(1-\cos(2k_l\hat z))}$ is the periodic lattice potential with lattice depth $U_0$, lattice wave number $k_l$ and the vertical spatial coordinate $\hat z$, and ${\hat H_g = - m_a g \hat z}$ represents the gravitational potential considered in first approximation as a linear potential, where $m_a$ is the atomic mass and $g$ the gravity acceleration.

\begin{figure}[ht]
    \begin{center}
    	\includegraphics[width=8.5 cm]{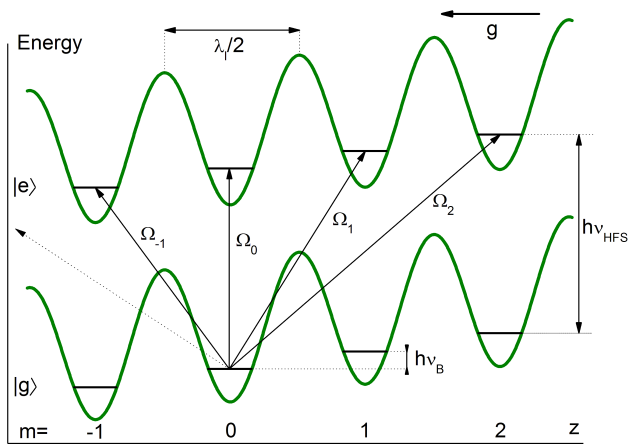}
		\end{center}
   	\caption{(Color online) Wannier-Stark ladder where $\nu_B$ is the Bloch frequency, $\nu_{HFS}$ the hyperfine transition between the states $\left|g\right\rangle=\left|5^{2}S_{1/2},F=1,m_{F}=0\right\rangle$ and $\left|e\right\rangle=\left|5^{2}S_{1/2},F=2,m_{F}=0\right\rangle$ of $^{87}$Rb, $m$ the quantum number corresponding to the lattice sites and ${\Omega_{\Delta m}}$ the coupling between the wells $m$ and ${m\pm\Delta m}$.}
    \label{Fig1_WSladder}
\end{figure}

Solving the time independent Schr\"odinger equation, one obtains the so-called Wannier-Stark ladder of states known from solid state physics~\cite{WannierStark1, WannierStark2} where the eigenstates $\left| W_m \right\rangle$ are approximately centered in the well labelled by the quantum number $m$. Each well is separated from its neighbour by an increment in potential energy from which the Bloch frequency $\nu_B$ is defined:
\begin{equation}
h \nu_B = m_a g \lambda_l/2,
\label{eq:NuB}
\end{equation}
where $\lambda_l$ is the lattice wavelength (see Fig.~\ref{Fig1_WSladder}).
 
With respect to Bloch states which are, in the absence of a linear force, delocalized all along the lattice due to its periodicity~\cite{BlochStates1, BlochStates2}, the spread of the atomic wavefunction $\left| W_m \right\rangle$ depends on the lattice depth. While being well localized in the well $m$ at high depth (${U_0 \gg 10}~E_r$), the wavefunction extends across a significant number of wells when reducing the depth below $5$ $E_r$~\cite{PRA_Messina}, where the recoil energy $E_r$ is defined by ${E_r/\hbar = \frac{\hbar k_l^2}{2m_a} \simeq 2\pi \times 8}$~kHz.

This delocalization allows resonant induced tunnelling between different lattice wells~\cite{PRL_Beaufils}. This tunnelling is realized with a two-photon Raman transition connecting the two hyperfine levels of the $^{87}$Rb ground state, $\left|g\right\rangle$ and $\left|e\right\rangle$, using counter-propagating vertical beams. This implies a momentum transfer of ${k_{eff} = k_{R1} + k_{R2} \approx 4\pi/\lambda_{Raman}}$, where \mbox{${\lambda_{Raman} \simeq 780}$~nm}.
For ${k_{eff} \approx k_l}$, the coupling between $\left| W_m, g \right\rangle$ to $\left| W_{m'}, e \right\rangle$ either in the same well (${m = m'}$) or in neighboring wells (${m \neq m'}$) is relatively large~\cite{PRA_Tackmann} and leads to Rabi oscillations with a Rabi frequency given by~\cite{PRA_Wolf_Fromopticallatticeclock}:
\begin{equation}
\Omega_{\Delta m}=\Omega_{U_{0}=0}\left\langle W_{m}\right|e^{-ik_{eff}\hat{z}}\left|W_{m+\Delta m}\right\rangle,
\label{eq:Omega}
\end{equation}
where ${\Omega_{U_{0}=0}}$ is the two-photon Rabi frequency in the absence of the lattice potential. ${\Omega_{\Delta m}}$ does not depend on the initial site $m$ but on the absolute value of the well separation ${\Delta m = m'-m}$ and on the lattice depth $U_0$. 

As Raman transitions provide labelling of the external via the internal atomic state~\cite{PLA_Borde}, a hyperfine selective detection allows for detecting the change in the external atomic state and thus the transport process in the lattice.

\section{Experimental set-up}

\begin{figure}[ht]
    \begin{center}
    	\includegraphics[width=8.5 cm]{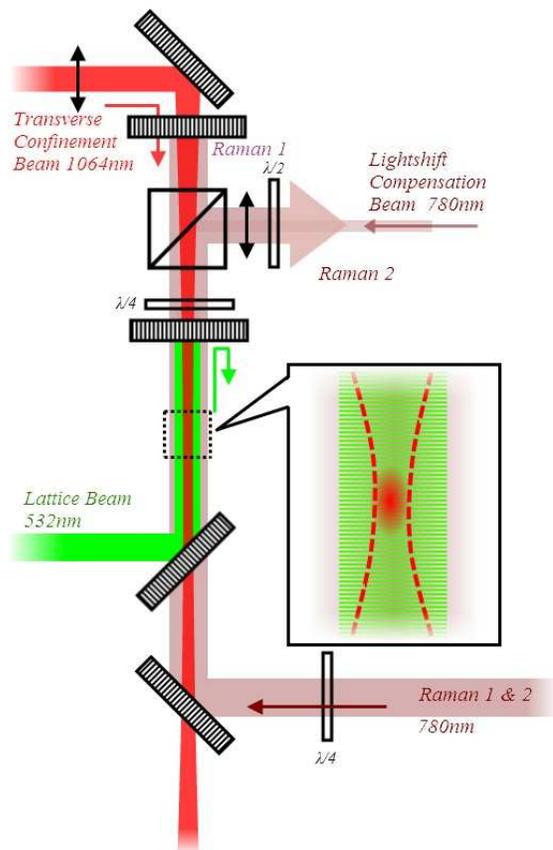}
		\end{center}
   	\caption{(Color online) Experimental setup. Laser beams for optical trapping and Raman transitions are superimposed using dichroic mirrors. To allow for counter-propagating transitions, one of the Raman beams is retro-reflected.}
    \label{Fig2_Setup}
\end{figure}

Our 1D-optical lattice is created by a vertically retro-reflected and single mode frequency doubled Nd:YVO$_{4}$ laser (${\lambda_{l} = 532}$~nm, $P = 8$~W, and $1/e^{2}$~radius of $600$~$\mu$m on the atoms). It is loaded from a 3D-Magneto-Optical Trap (MOT) containing ${3 \times 10^{7}}$~atoms cooled down to $2$~$\mu$K with $100$-MHz-detuned molasses. Superimposed on the lattice, a red detuned progressive wave from an Yb fiber laser (${\lambda = 1064}$~nm, ${P \simeq 2}$~W, and ${1/e^{2}}$~radius of $180$~$\mu$m on the atoms) is added to provide a transversal confinement (see Fig.~\ref{Fig2_Setup}). In this mixed-trap, a few $10^{4}$~atoms are distributed in about $4000$~adjacent Wannier-Stark states (along about $1$~mm) in the fundamental lattice band. The typical trapping lifetime is about $1$~s. As the transverse dipole trap induces a large differential light shift on the hyperfine states of the atoms, a compensating beam well mode matched on it is used to lead to its cancellation; the implementation details of this compensating beam can be found in~\cite{PRL_Beaufils}. In addition to that, the circularly polarized Raman beams are carefully superimposed on the transverse dipole trap to avoid couplings between the transverse states induced by the horizontal component of the Raman effective momentum $k_{eff}$. The Raman lasers are red detuned from the $^{87}$Rb D2 line by about $3$~GHz, with a maximum power of ${P_{max} = 1.7}$~mW distributed over a $1/e^{2}$~radius of $5$~mm on the atoms. This configuration allows us to confine the atoms at the center of the 1D-lattice in order to prevent for quick losses of atoms while reducing variations of the lattice depth $U_{0}$ across the cloud and thus coupling inhomogeneities.

After loading this mixed trap, an internal state preparation is performed to transfer all the atoms in the $\left|5^{2}S_{1/2},F=1,m_{F}=0\right\rangle$ state to reduce the sensitivity to stray magnetic fields. The atoms accumulated during the MOT in all the Zeeman sublevels of $\left|5^{2}S_{1/2},F=2\right\rangle$ are depumped to $\left|5^{2}S_{1/2},F=1\right\rangle$ and then optically pumped (with $95 \%$ efficiency) 
to the $\left|5^{2}S_{1/2},F=1,m_{F}=0\right\rangle$ Zeeman sublevel. At this stage, the atoms are ready to be interrogated by Raman or microwave pulses. Finally, a time-of-flight fluorescence state selective detection is used to measure the populations in the two hyperfine states ($N_g$, $N_e$) after the release of the atoms from the trap. From the population measurement, the transition probability from $\left|g\right\rangle$ to $\left|e\right\rangle$ is then computed as ${P_e = \frac{N_e}{N_e + N_g}}$. Using Raman transitions as a probe tool, an almost Fourier-limited spectroscopy of the Wannier-Stark states is possible~\cite{PRA_Tackmann}, where tunnelling is induced between neighbouring wells when the Raman frequency difference fulfills the resonance condition ${\nu_{R1}- \nu_{R2} = \nu_{HFS} + \Delta m \times \nu_B}$. We have demonstrated sub-Hz resolution of our Wannier-Stark resonance, with FWHM of $0.7$~Hz using a Raman $\pi$-pulse of ${\tau_{R} = 1.4}$~s duration.

\section{Ramsey-Raman interferometer}

To perform a high resolution measurement of the transition frequency between two coupled Wannier-Stark states, we apply a Ramsey-type sequence of two Raman $\pi/2$-pulses separated by a time ${T_{R}}$, for which the frequency resolution is ${\Delta \nu \sim 1/(2T_{R})}$. Those pulses prepare, let evolve and recombine a superposition of external (and also internal) states, where partial atomic wavepackets are centered in two different wells. We have previously demonstrated~\cite{PRL_Beaufils} a maximum coherence time of the order of $1$ s and efficient spatial separation of up to ${\Delta m = \pm7}$ wells for a trap depth of ${1.8~E_r}$.

The interferometer phase is given by:
\begin{equation}
\Delta \phi = 2 \pi \left( \nu_R - \left(\tilde{\nu}_{HFS} + \Delta m \times \nu_B \right)\right) \tilde{T}_{R},
\label{eq:RamRamPhase}
\end{equation}
where $\Delta \phi$ is the atomic phase difference at the end of the interferometer, ${\tilde{\nu}_{HFS}}$ is the hyperfine transition frequency, perturbed by Raman or dipole trap light shifts and quadratic Zeeman effect, and ${\nu_R = \nu_{R1} - \nu_{R2}}$ is the frequency difference between the two Raman beams. ${\tilde{T}_{R}}$ is the effective Ramsey time, which, in the case of finite duration Raman pulses, depends on the Raman pulse duration $\tau_R$. The effective Ramsey time is given by ${\tilde{T}_{R} = T_{R} + \frac{4\tau_R}{\pi}}$; this expression can be easily derived using the sensitivity function of the interferometer~\cite{Proc_Dick}.

\begin{figure*}[ht]
    \begin{center}
    	\includegraphics[width=18 cm]{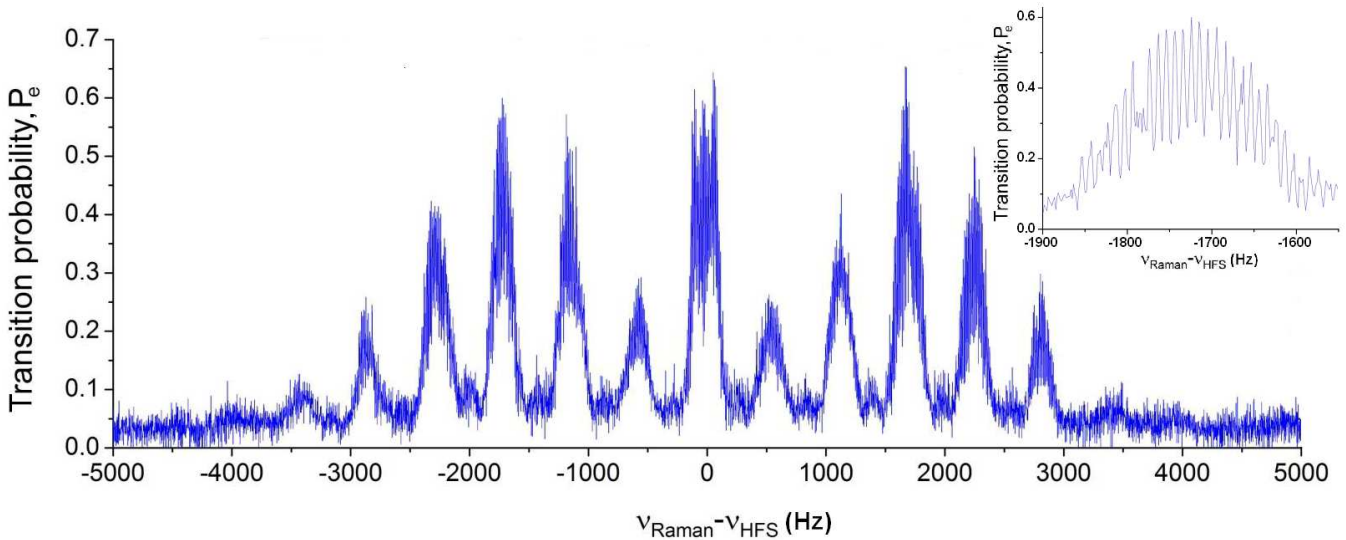}
   	\caption{Ramsey-Raman fringes for a lattice depth ${U_0 = 3.9~E_r}$ showing evidence of transitions between up to $6$ neighboring lattice sites. Rabi envelopes are separated by the Bloch frequency ${\nu_B = 568.5}$~Hz and contain interference fringes separated by ${\Delta \nu = 1/\tilde{T}_{R}}$. Here ${\tau_{R} = 5}$~ms and ${T_{R} = 100}$~ms. The intensities of the Raman lasers and the lattice depth were set to optimize the contrast of the ${\Delta m = \pm 3}$ fringes. Insert: zoom on the ${\Delta m = -3}$ transition.}
    \label{Fig3_Interferences}
    \end{center}
\end{figure*}

Fig.~\ref{Fig3_Interferences} displays the interferometer fringe pattern, recorded by scanning the frequency difference between the two Raman lasers across the numerous resonances corresponding to different $\Delta m$ transitions. In each Rabi envelope we observe a set of Ramsey fringes (see insert in Fig.~\ref{Fig3_Interferences}).

\begin{figure}[ht]
    \begin{center}
    	\includegraphics[width=8.5 cm]{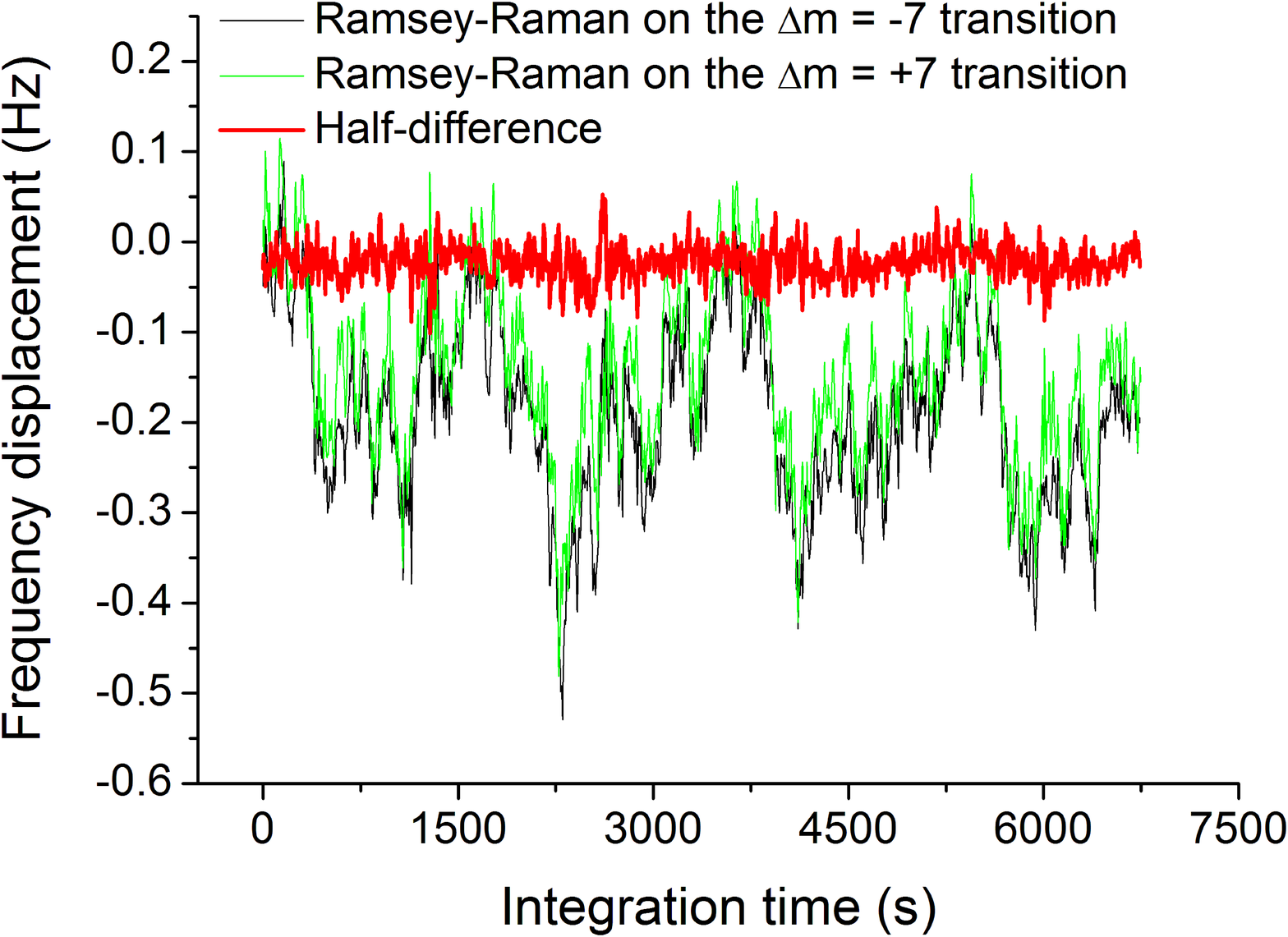}
   	\caption{(Color online) Ramsey-Raman frequency fluctuations measured on the ${\Delta m = -7}$ (black line) and ${\Delta m = +7}$ (green line) transitions for experimental parameters ${\tau_{R} = 120}$ ms and ${T_{R} = 850}$ ms. The half-difference (red thick line) cancels the hyperfine transition clock frequency drift.}
    \label{Fig4_Half-diff_temp}
    \end{center}
\end{figure}

The position of the central fringe is then measured using a computer controlled frequency lock. In practice, we alternate transition probability measurements on both sides of the central fringe from which we extract an error signal to determine the frequency offset with respect to the position of the central fringe. We thus use consecutive transition probability measurement to stir $\nu_R$ onto the central fringe frequency, realizing a numerical integrator. In comparison to a fringe fitting measurement, the half-fringe transition probability measurement contains only the most sensitive points, which have steeper slope for the measurement of the fringe position, which thus leads to maximum sensitivity to phase and frequency fluctuations. Fig.~\ref{Fig4_Half-diff_temp} displays in black the frequency fluctuations on the measurement of the ${\Delta m = -7}$ transition. For this measurement the experimental parameters are ${\tau_{R} = 120}$~ms and ${T_{R} = 850}$~ms, and the contrast of the interferometer fringes is about $50\%$. We observe fluctuations up to $400$~mHz peak-to-peak, well correlated with temperature fluctuations of the lab. Interleaving between two ${\Delta m = -7}$ measurement, we measure frequency fluctuations on the ${\Delta m = +7}$ intersite transition with the same frequency lock (displayed in green in Fig.~\ref{Fig4_Half-diff_temp}) which are well correlated with the ${\Delta m = -7}$ measurement. This illustrates the fact that these frequency fluctuations are dominated by the fluctuations of the effective hyperfine transition frequency ${\tilde{\nu}_{HFS}}$. Taking the half-difference of those two ${\pm \Delta m}$ frequencies, we obtain a measurement of ${7 \times \nu_B}$ free from clock related frequency shifts. The efficiency of this rejection is illustrated by the significant reduction of the fluctuations amplitude observed on the red thick trace of the Fig.~\ref{Fig4_Half-diff_temp}.

\begin{figure}[ht]
    \begin{center}
    	\includegraphics[width=8.5 cm]{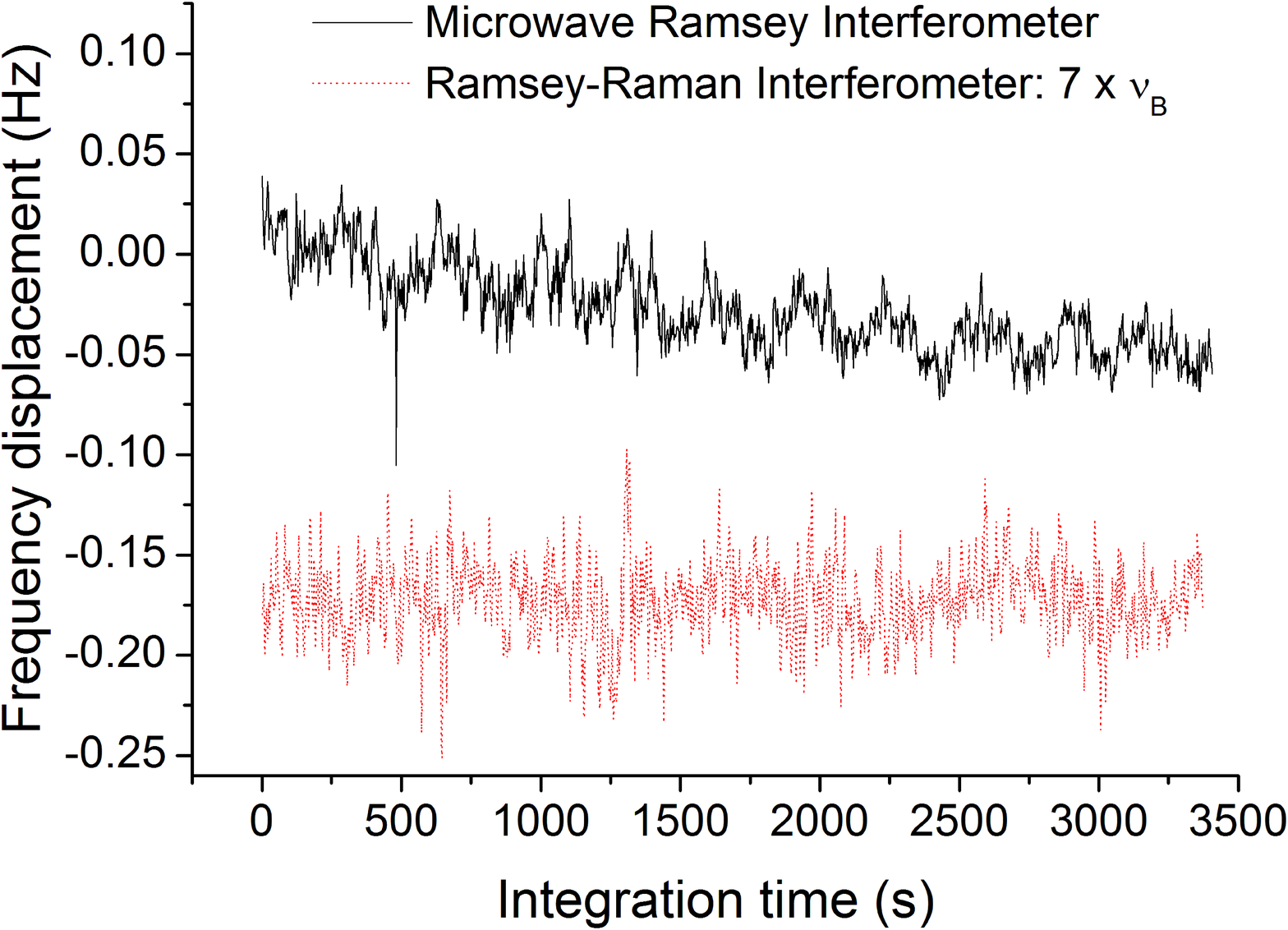}
   	\caption{(Color online) Frequency fluctuations of $\nu_{HFS}$ measured with a microwave Ramsey interferometer for experimental parameters ${\tau_{MW} = 0.5}$~ms and ${T_{R} = 1 476}$~ms. Fluctuations of the half-difference of the Ramsey-Raman interferometer on the ${\Delta m = \pm 7}$ transitions are displayed as a dashed red line for comparison; an artificial offset is present for clarity on the graph.}
    \label{Fig5_ClockFreq_Temp}
    \end{center}
\end{figure}

To discriminate the relative influence of the different sources of the frequency fluctuations, we compare our Ramsey-Raman to a standard Ramsey microwave interferometer realized using two $\pi/2$-microwave transitions separated by a Ramsey time $T_{R}$. In the latter case, microwave pulses allow for coupling between hyperfine states only, without any change in the well index due to the much smaller momentum transfer of the microwave photon compared to the lattice momentum. The position of the central fringe corresponds to the hyperfine transition frequency perturbed by the quadratic Zeeman effect, the differential light shifts of the trapping lasers and the compensating beam. As our microwave synthesizer has a too large minimum resolution of $0.1$~Hz in frequency, we do not use the numerical integrator scheme detailed above. Instead we use transition probability measurements at mid-fringes (alternatively on the right and left side of the fringe), from which we derive the frequency fluctuations of the central fringe of the Ramsey microwave pattern on the ${\Delta m = 0}$ transition. For a microwave pulse of ${\tau_{MW} = 0.5}$~ms and a Ramsey time ${T_{R} = 400}$~ms, we find fluctuations of about $100$~mHz (see Fig.~\ref{Fig5_ClockFreq_Temp}). This is $4$ times less than for the Ramsey-Raman interferometer but consistent with the measured power fluctuations of the involved lasers and magnetic field listed in Table~\ref{Tab1_FluctLSandB}.

\begin{table}
\caption{Summary of the measurements of the different effects inducing differential frequency shifts and their relative fluctuations.\label{Tab1_FluctLSandB}}
\begin{ruledtabular}
\begin{tabular}[t]{lll}
Transverse dipole trap & $-4$~Hz & $<1\%$ \\
Compensating beam & $+4$~Hz & $1\%$ \\
Lattice beam & $+0.4$~Hz & a few $\%$ \\
Quadratic Zeeman effect & $+1.5$~Hz & $10^{-3}$
\end{tabular}
\end{ruledtabular}
\end{table}

The amplitude and the stability of the light shift induced by the Raman lasers was then studied using microwave spectroscopy. For this purpose, we shine in an off-resonant Raman light field with an intensity corresponding to $\Omega\tau\approx 6.5 \pi$ during the application of a microwave $\pi$-pulse duration of ${\tau = 120}$~ms. For comparison with a resonant case for a Raman pulse with no microwave pulse, this Raman power would thus drive a $\pi$-pulse in a relatively short time of $18.5$~ms which corresponds to a Rabi frequency of ${\Omega/2\pi = 27}$~Hz on the ${\Delta m = -3}$ transition at $4~E_r$. This intensity setting enhances the effect of the AC Stark shift of the Raman beams, and the detuning from resonance prevents any coupling between the states.

\begin{figure}[ht]
    \begin{center}
    	\includegraphics[width=8.5 cm]{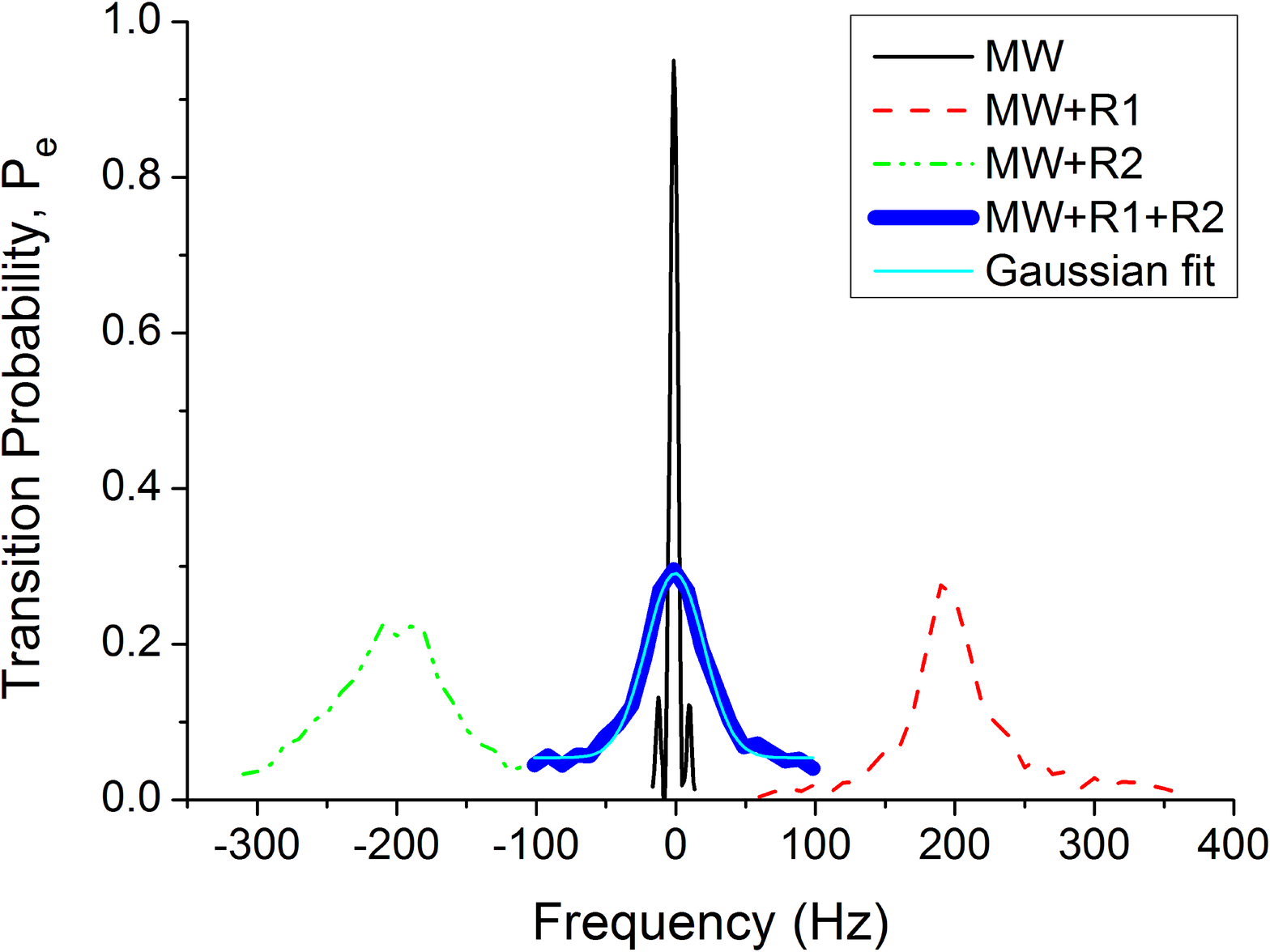}
   	\caption{(Color online) Differential light shift of the different Raman beams which is characterized by its mean light shift $\nu_{RDLS}$ and its frequency inhomogeneities $\sigma_{RDLS}$, measured by microwave spectroscopy with/without Raman beam out of resonance. Measurements are made for Raman pulses inducing a Rabi frequency of $27$ Hz on the ${\Delta m = -3}$ transition. From a Gaussian fit on the microwave spectroscopy with the two Raman beams, frequency inhomogeneities are determined to be $\sigma_{RDLS} = 20$~Hz.}
    \label{Fig6_RamanLS}
    \end{center}
\end{figure}

Fig.~\ref{Fig6_RamanLS} displays the scans of the transition probability for a microwave pulse alone (black solid trace), with one of the Raman laser (green dashed two dotted trace), with the other one (red dashed trace), and with both (blue thick solid trace). As the two Raman beams have opposite light shift due to their red detuning of $3$ GHz from resonance, their power ratio can be chosen so that the mean Raman differential lighshift ($\nu_{RDLS}$) is cancelled when they are both present. Despite adjusting this power ratio, we find on Fig.~\ref{Fig6_RamanLS} rather large inhomogeneities of the resonance frequency. A Gaussian fit (light blue thin solid line) of the resonance curve gives a standard deviation of ${\sigma_{RDLS} = 20}$~Hz, which is typically of the same order of magnitude as the half of the Rabi frequency. We attribute this width to inhomogeneous longitudinal and transverse intensity distributions of the Raman beams in the mixed trap.

In addition, we have measured the fluctuations of the mean value of this Raman differential light shift using microwave spectroscopy. We found fluctuations of $\nu_{RDLS}$ of $2$~Hz peak-to-peak with Raman laser intensities adjusted to drive a $\pi$-pulse in $120$~ms on the transition ${\Delta m = -3}$. The corresponding measured light shifts of individual Raman lasers are then about ${\pm 30}$~Hz. The measured fluctuations are found to be significantly larger than expected from Raman laser intensity fluctuations (of about $1\%$, which would imply fluctuations of about $420$~mHz assuming uncorrelated fluctuations). We thus attribute these fluctuations to fluctuations of the position of the trapped atoms in the inhomogeneous transverse profile of the Raman lasers. 

The resulting displacement of the central fringe is given by the Raman light shift filtered by the sensitivity function of the interferometer~\cite{Proc_Dick}:
\begin{equation}
\delta \nu_{R} = \frac{4 \tau_R}{\pi \tilde{T}_{R}} \nu_{RDLS},
\label{eq:RamanLS}
\end{equation}
where ${\delta \nu_R}$ is the shift in frequency of the central fringe of the interferometer pattern.

Taking into account that the Raman coupling is transition dependent (see Eq.~\ref{eq:Omega}) leading to a ratio ${\Omega_{\pm 3}/\Omega_{\pm 7} = 1.5}$ at $4~E_r$ for $\Delta m = \pm 3$ and at $1.8~E_r$ for $\Delta m = \pm 7$, we calculate the impact of these measured Raman differential light shift fluctuations of $2$~Hz peak-to-peak on the Ramsey-Raman interferometer frequency fluctuations to be on the order of $400$~mHz peak-to-peak in good agreement with the measurement in Fig.~\ref{Fig4_Half-diff_temp}. We thus attribute these fluctuations to this latter effect, the mean Raman differential light shift fluctuations.

\section{Sensitivity}

\begin{figure}[ht]
    \begin{center}
    	\includegraphics[width=8.5 cm]{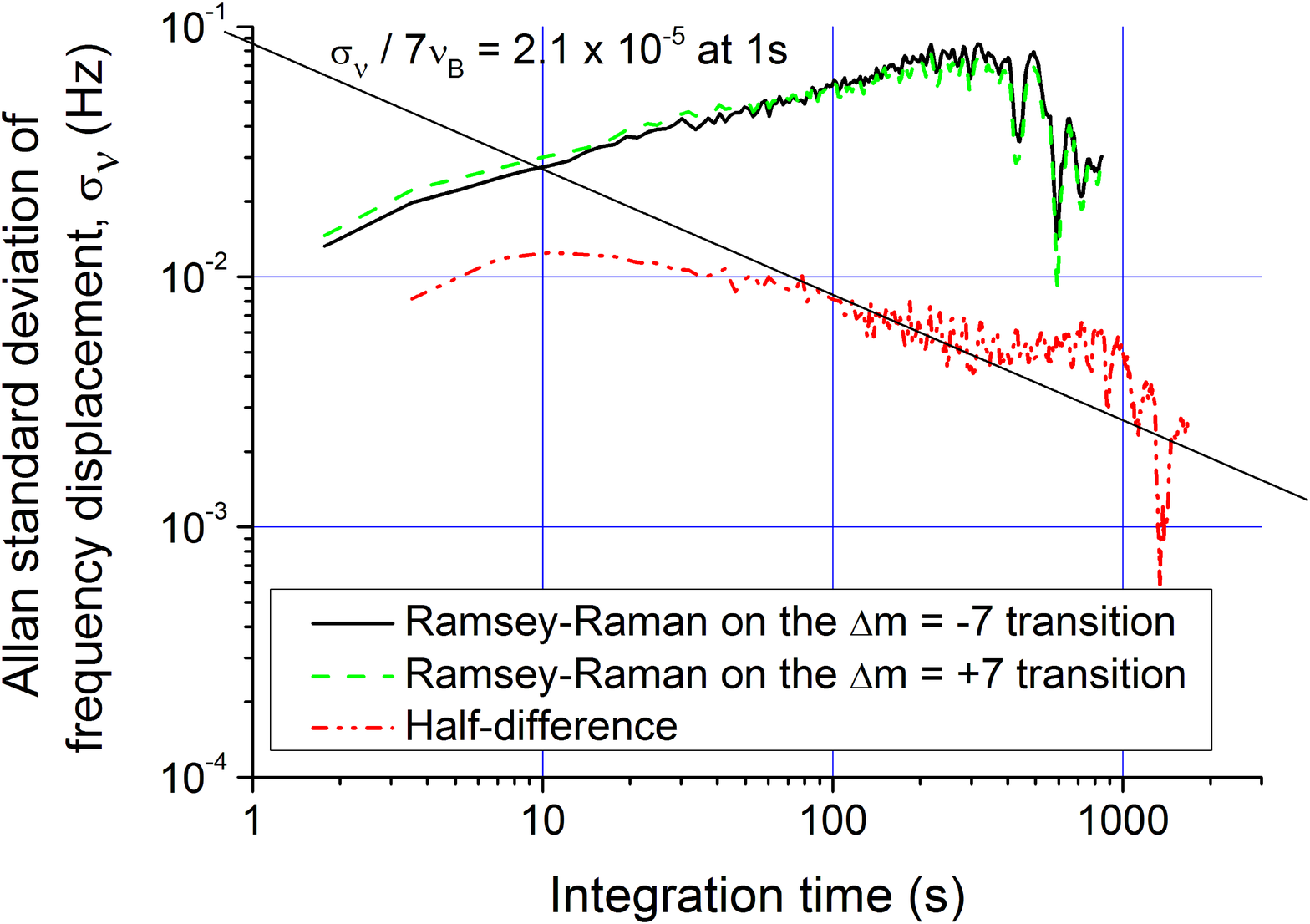}
   	\caption{(Color online) Allan standard deviation of the frequency fluctuations of $\nu_B$ on the ${\Delta m = +7}$, ${\Delta m = -7}$ transitions measured with a Ramsey-Raman interferometer and the computed half-difference of those two measurements cancelling the hyperfine structure component, for experimental parameters ${\tau_{R} = 120}$~ms and ${T_{R} = 850}$~ms.}
    \label{Fig7_Half-diff_AllanStd}
    \end{center}
\end{figure}

Computing the Allan standard deviation of those frequency fluctuations of the half-difference on the ${\Delta m = +7}$ and ${\Delta m = -7}$ transitions (see Fig.~\ref{Fig7_Half-diff_AllanStd}), we obtain an equivalent short-term sensitivity of ${8.8 \times 10^{-2}}$~Hz at $1$~s which corresponds to a relative sensitivity of ${\frac{\delta \nu}{\nu}=\frac{\sigma_{\nu}}{7\nu_B}=2.2 \times 10^{-5}}$ at $1$~s. It is important to notice that the short-term sensitivity on the Allan standard deviation measurements in the Fig.~\ref{Fig7_Half-diff_AllanStd} is filtered by the time constant of the numerical integrator used to lock the Raman frequency difference on the central fringe of the interferometer.

\begin{figure}[ht]
    \begin{center}
    	\includegraphics[width=8.5 cm]{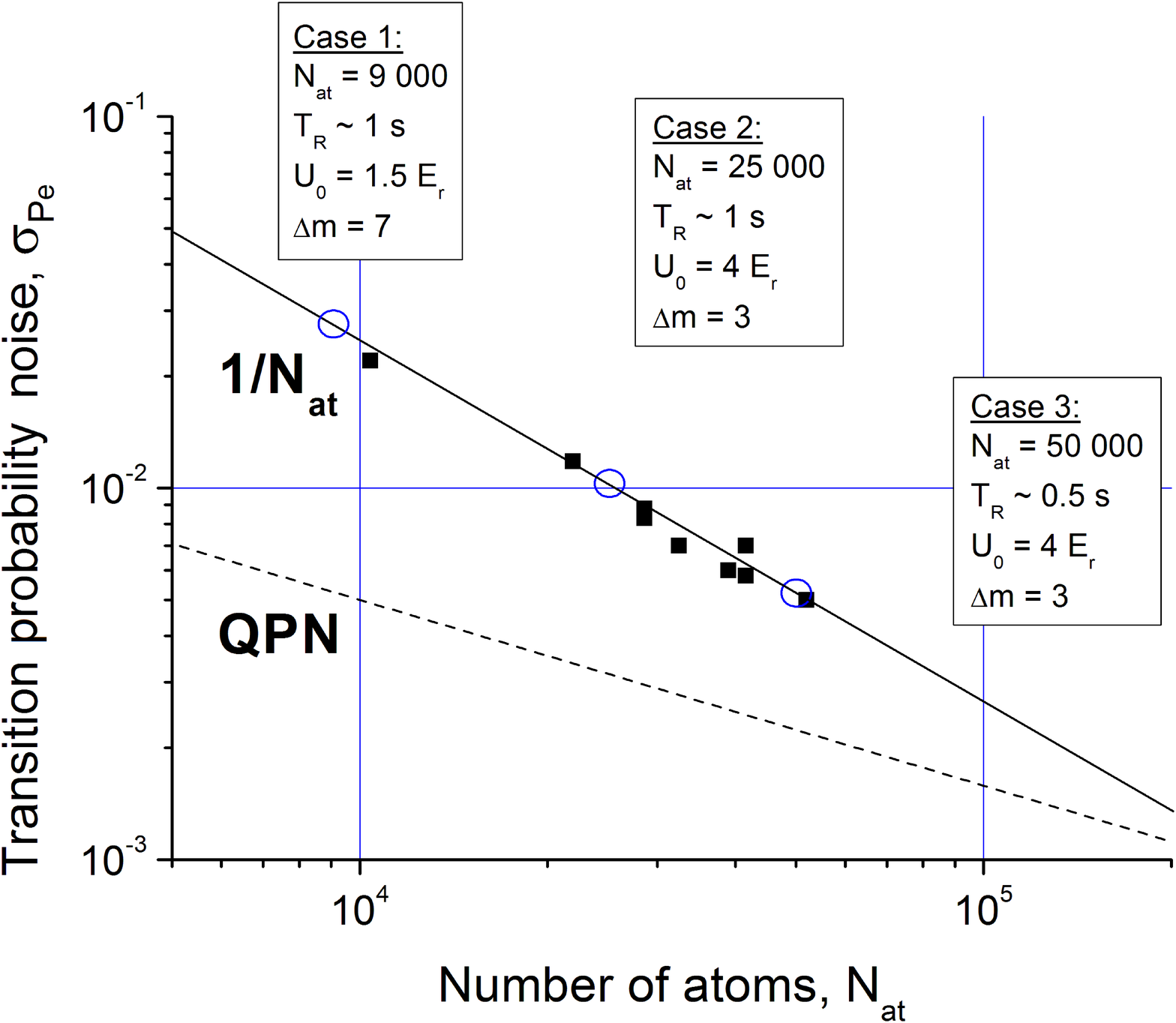}
   	\caption{Short-term detection noise as a function of the number of measured atoms. $\sigma_{P_e}$ corresponds to the $1$-shot Allan standard deviation of $P_e$. Each circle corresponds to a different set of experimental parameters. The quantum projection noise is illustrated with the dashed line.}
    \label{Fig8_Short-term_noise}
    \end{center}
\end{figure}

Remarkably, other sets of experimental parameters lead to the same relative sensitivity (see Fig.~\ref{Fig8_Short-term_noise}). Indeed the short-term limitation in this measurement is mainly due to the detection noise which is given by:
\begin{equation}
\frac{\delta \nu}{\nu}=\frac{\sigma_{\nu}}{\Delta m \times \nu_B}=\frac{\sigma_{P_e}}{\pi C T_R \Delta m \times \nu_B},
\label{eq:Det_Noise}
\end{equation}
with $\sigma_{P_e}$ the transition probability noise and $C$ the contrast of the interference fringes. As the detection noise is limited by the electronic noise of the photodiode used to collect the fluorescence from the atoms in the states $\left|g\right\rangle$ and $\left|e\right\rangle$, it can be written with two balanced channels ${\sigma_{P_e} = \frac{\sqrt{2}\sigma_{elec}}{2(N_g + N_e)}}$, where ${\sigma_{elec} = 325}$~atoms for the photodiodes and transimpedance circuits we use.

To illustrate the impact of the choice of the measurement parameters, Fig.~\ref{Fig8_Short-term_noise} shows three sets of optimized parameters that are all detection noise limited but finally correspond to the same sensitivity on the Bloch frequency measurements, due to the interplay between experimental parameters. For example, the maximal spatial separation ${\Delta m}$ is linked to the spatial spread of the atomic wavefunction, and increases when decreasing lattice depth~\cite{PRA_Messina}. But decreasing the lattice depth also decreases the number of trapped atoms ${N_{at} = N_g + N_e}$, so that the gain in the intrinsic phase sensitivity (proportional to ${\Delta m}$) is compensated by the increase in detection noise. Alternatively, increasing the trapping time in order to increase the Ramsey time $T_{R}$ also decreases the number of detected atoms. There is thus a trade-off between lattice wells separation $\Delta m$, interferometer time $T$ and corresponding number of atoms $N_{at}$ for the used trapping time. The remaining independent parameters that limit the sensitivity are the electronic noise of the detection and the contrast of the interferometer, which is mainly limited by the efficiency of the Raman coupling ($80\%$~at most) due to inhomogeneities of resonance frequency and coupling.

\begin{figure}[ht]
    \begin{center}
    	\includegraphics[width=8.5 cm]{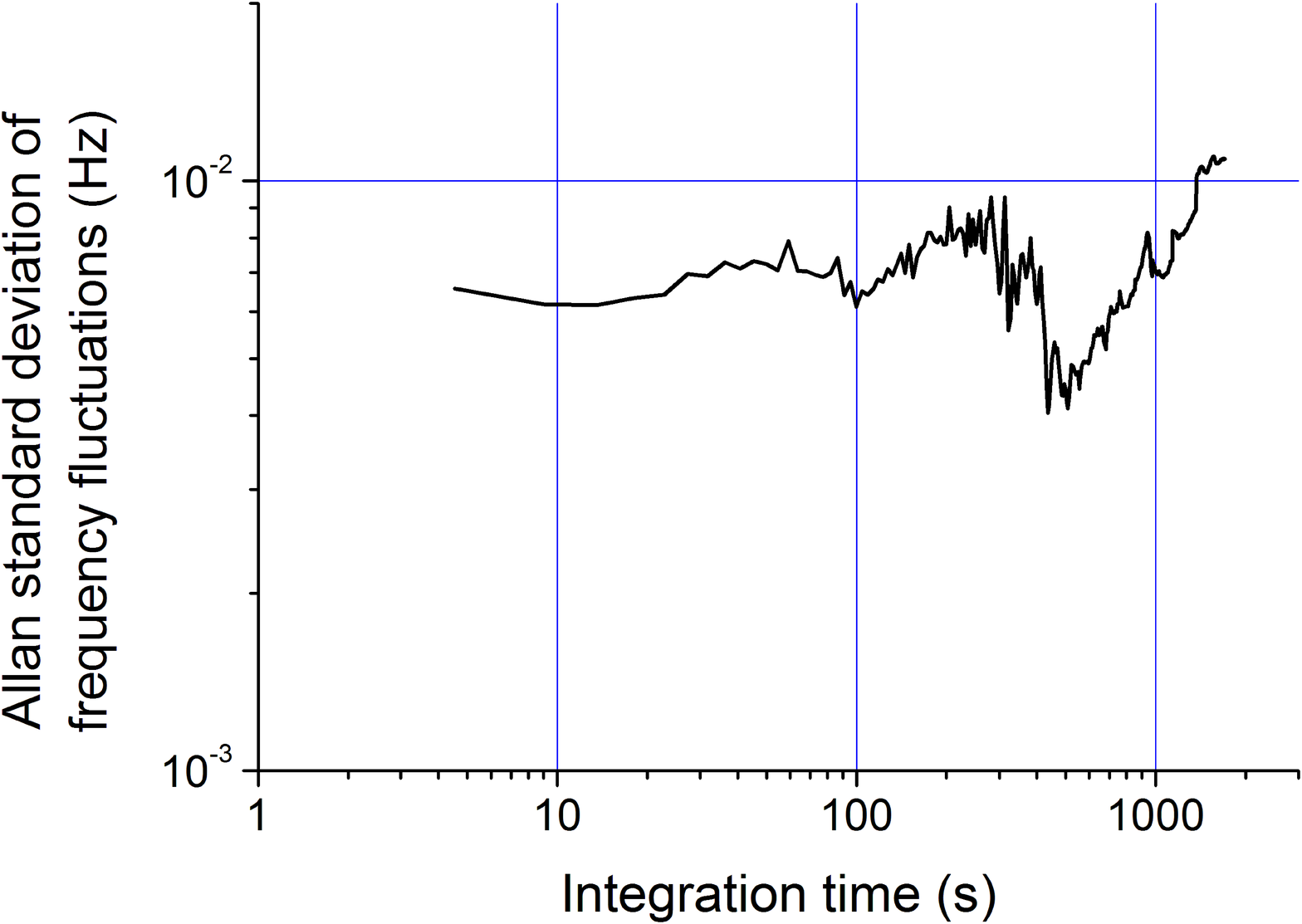}
   	\caption{Allan standard deviation of the frequency fluctuations of $\nu_{HFS}$ measured with a microwave Ramsey interferometer of ${\tau_{MW} = 0.5}$~ms and ${T_{R} = 1 476}$~ms.}
    \label{Fig9_ClockFreq_Astd}
    \end{center}
\end{figure}

For comparison, we show in Fig.~\ref{Fig9_ClockFreq_Astd} the Allan standard deviation of the clock frequency fluctuations measured using the standard Ramsey microwave interferometer described above. We obtain a better frequency sensitivity of ${\sigma_{\nu} = 6.6 \times 10^{-3}}$~Hz for a two samples measurement, corresponding to a measurement time of $4.4$~s. This is due to a longer interrogation time of ${T_{R} \simeq 1.5}$~s, a better contrast of $80\%$, and a larger number of atoms due to a higher depth. This corresponds to a relative sensitivity on the clock frequency of ${\sigma_{\nu}/\nu_{HFS} = 9.6 \times 10^{-13}}$ at $4.4$~s. With respect to Fig.~\ref{Fig7_Half-diff_AllanStd}, the short-term sensitivity is, here, not filtered by the time constant of the integrator.

\section{Systematics}

We now focus on the main systematic effect in our measurement which is related to the intensity gradient of the transverse dipole trap beam along the lattice axis. This intensity gradient arises from the focusing of the transversal trapping beam on the atoms. This leads to a parasitic force if atoms are not precisely located at the waist position, which shifts the value of the Bloch frequency. 
Considering this dipole trap beam as a pure TEM00 mode progressive wave, we calculate a relative bias on the Bloch frequency as a function of the distance between the atoms and the beam waist of ${1.1 \times 10^{-5}}$/mm for a power of $2$~W, when the atoms are (transversally) at the center of the transverse dipole trap beam. The effect of the intensity gradient of the lattice beam is much smaller due to its larger waist and smaller depth: we find a relative shift of ${3 \times 10^{-9}}$/mm at $8$~W of power.

\begin{figure}[ht]
    \begin{center}
    	\includegraphics[width=8.5 cm]{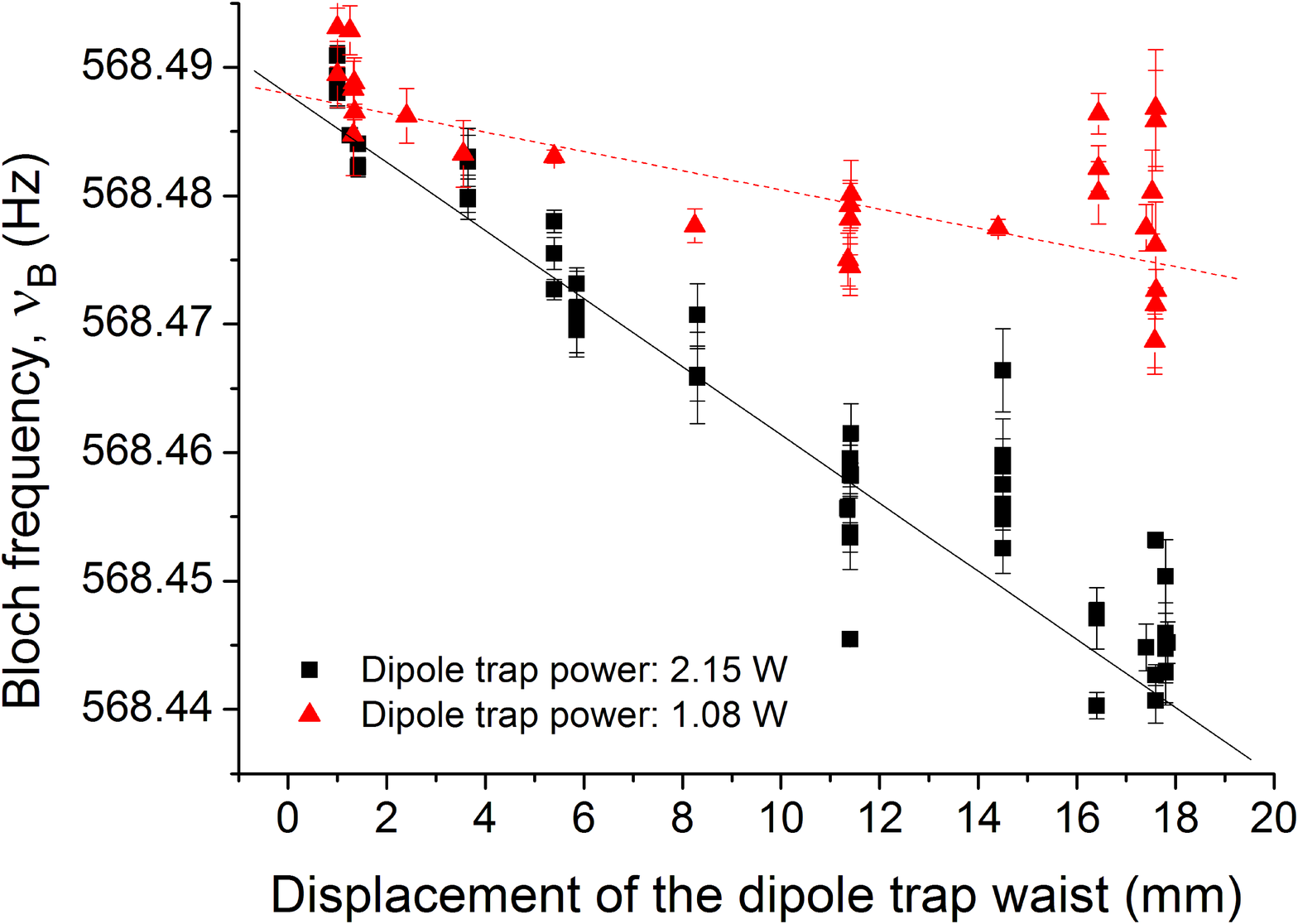}
   	\caption{(Color online) Systematic effects on the absolute frequency measurement of $\nu_B$ due to the intensity gradient of the dipole trap as a function of the distance to the dipole trap waist. Solid and dashed lines are linear fits to the data.}
    \label{Fig10_Systematics_IR}
    \end{center}
\end{figure}

Fig.~\ref{Fig10_Systematics_IR} displays measurements of the Bloch frequency as a function of the change in position of the atoms with respect to the transverse dipole trap beam waist position for two different laser power values. In order to change the relative position of the atoms with respect to the waist, we shift the position of the dipole trap beam focusing lens over about $2$~cm. Despite a relatively large scatter in the data, we find shifts that roughly scale linearly versus position, as expected. The position of the lens for which the atoms are at the waist should correspond to the position where the two linear fits cross, and the corresponding value of the Bloch frequency is found to be $568.488(1)$~Hz. For the largest power, of $2.15$~W, we find a shift of $-2.6$~mHz/mm, which corresponds to a relative shift of ${4.5 \times 10^{-6}}$/mm, significantly smaller than expected. We attribute that to the averaging of the longitudinal force over the radial direction.
Moreover, the ratio of the slopes is found to be $3.5$, much different from the ratio of the two laser powers of $2$. This may be due to different radial atomic density distributions at different laser powers and thus different averaging of the force over the radial direction. A detailed study of this effect would require a better reproducibility as well as systematic measurements of the atomic position and density distribution in the transverse dipole trap.

The value of the gravity acceleration can be derived from the value of the Bloch frequency, provided the laser wavelength is known (see Eq.~\ref{eq:NuB}). For that purpose, the lattice laser is locked on an Iodine line using a frequency modulation transfer spectroscopy. This line is the hyperfine transition $a1$ of the 1116/P(52)32-0 line at $532.195951(1)$~nm~\cite{APB_Iodinelines}. The relative uncertainty in the lattice laser frequency is estimated to be about ${\delta\nu/\nu = 2 \times 10^{-9}}$. Using the precise value of $h/m_a$ from \cite{PRL_Bouchendira} and the value mentioned above for the lattice laser wavelength (subtracting twice $80$~MHz due to a double pass AOM used for the spectroscopy), we derive $g = 9.80892(2)$~m/s$^2$. This value differs significantly from a previous measurement obtained in the lab with a free falling gravimeter~\cite{Metrologia_Merlet} of $g = 9.8092758(1)$~m/s$^2$. Although the source of this error remains to be identified, it does not arise from imperfect vertical alignment. Direct measurements of the Bloch frequency versus the tilt of the whole experimental setup were performed and residual misalignment is estimated to be lower than $1.7$~mrad, which corresponds to a maximum relative error of ${1.4 \times 10^{-6}}$.

\section{Accordion interferometer}

\begin{figure}[ht]
    \begin{center}
        \includegraphics[width=8.5 cm]{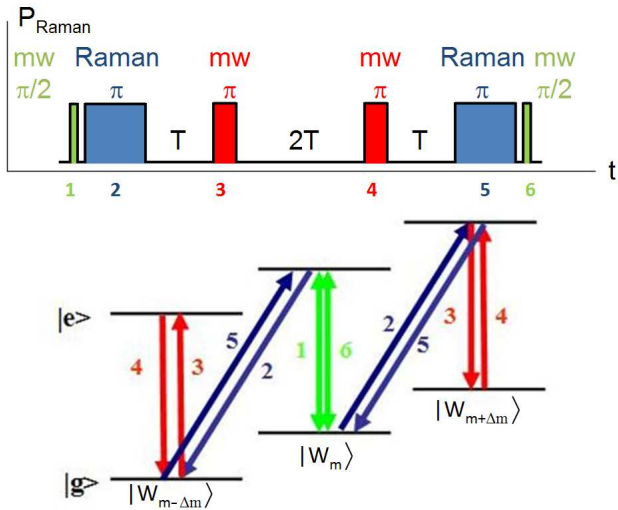}
    \caption{(Color online) Basic scheme of the accordion interferometer.}
    \label{Fig11_AccordeonInterferometer}
    \end{center}
\end{figure}

In the Ramsey-Raman interferometer measurements, the efficiency of the rejection of phase fluctuations induced by clock related shifts is limited by the switching time between the ${\pm \Delta m}$ measurements giving by the cycle time, ${\tau_C = 1.7}$~s. Fast fluctuations occurring on a time scale shorter or on the order of $\tau_C$ are not suppressed.
In order to improve the efficiency of this suppression, we have implemented the symmetric interferometer recently proposed in~\cite{PRA_Sophie}, which realizes an instantaneous measurement of ${-\Delta m \times \nu_B}$ and ${+ \Delta m \times \nu_B}$ at a time. The pulse sequence of this interferometer (which we refer to as the accordion interferometer) is displayed in Fig.~\ref{Fig11_AccordeonInterferometer}.

It is similar to other interferometer configurations where the use of additionnal (optical) $\pi$-pulses allows rejecting
undesired contributions in the interferometer phase such as clock frequency in free-fall inertial sensor~\cite{PLA_Borde, PRL_Kasevich_1991}, or linear acceleration in a symmetric $4$-pulses gyrometer~\cite{PhysB_Clauser, PRA_Landragin, PRL_Kasevich_Gyro} and in a multiple loop gradiometer~\cite{PRA_Kasevich_Gradio}.

This interferometer is made up of a composite sequence of microwave and Raman pulses. The use of short microwave $\pi$-pulses (of $1$~ms duration) is motivated by their much higher transfer efficiency between the hyperfine states of the same well (about $99\%$) when compared to Raman pulses.

Within this sequence, a first microwave $\pi/2$-pulse thus prepares an internal state superposition of the atomic wavefunction in the same well in the states $\left|g\right\rangle$ and $\left|e\right\rangle$. A Raman $\pi$-pulse changes their external state with a transition from the $\left|g\right\rangle$ state to the ${+ \Delta m}$ well, whereas the $\left|e\right\rangle$ state is transfered to the ${- \Delta m}$. A microwave $\pi$-pulse follows after a time $T$ to symmetrize the interferometer so that the atoms spend the same amount of time in the two hyperfine states. After a time $2T$, this half-sequence is reversed with a new microwave $\pi$-pulse, followed after a time $T$ by another Raman $\pi$-pulse. Finally a last microwave $\pi/2$-pulse recombines the atoms in the initial well after a separation of ${2 \times \Delta m = 3.72}$~$\mu$m (for ${\Delta m = 7}$).

With this symmetric interferometer, the (static) contribution from the hyperfine frequency shifts is in principle cancelled without the need of any alternating sequence. Indeed the phase difference at the output of the interferometer is independent of $\nu_{HFS}$:
\begin{eqnarray}
\Delta \phi & = & \phi_1 - 2\phi_2 + 2\phi_3 - 2\phi_4 + 2\phi_5 - \phi_6 \nonumber\\
& = & 16\pi \left(\nu_R - \Delta m \times \nu_B - \nu_{MW}\right) T,
\end{eqnarray}
where $\nu_{MW}$ is the microwave pulse frequency. For equal total interrogation times (taking into account that ${T_R \simeq 4T}$), the sensitivity on $\nu_B$ is here enhanced by a factor of two with respect to the Ramsey-Raman interferometer, owing to the increased separation between partial wavepackets.

\begin{figure}[ht]
    \begin{center}
        \includegraphics[width=8.5 cm]{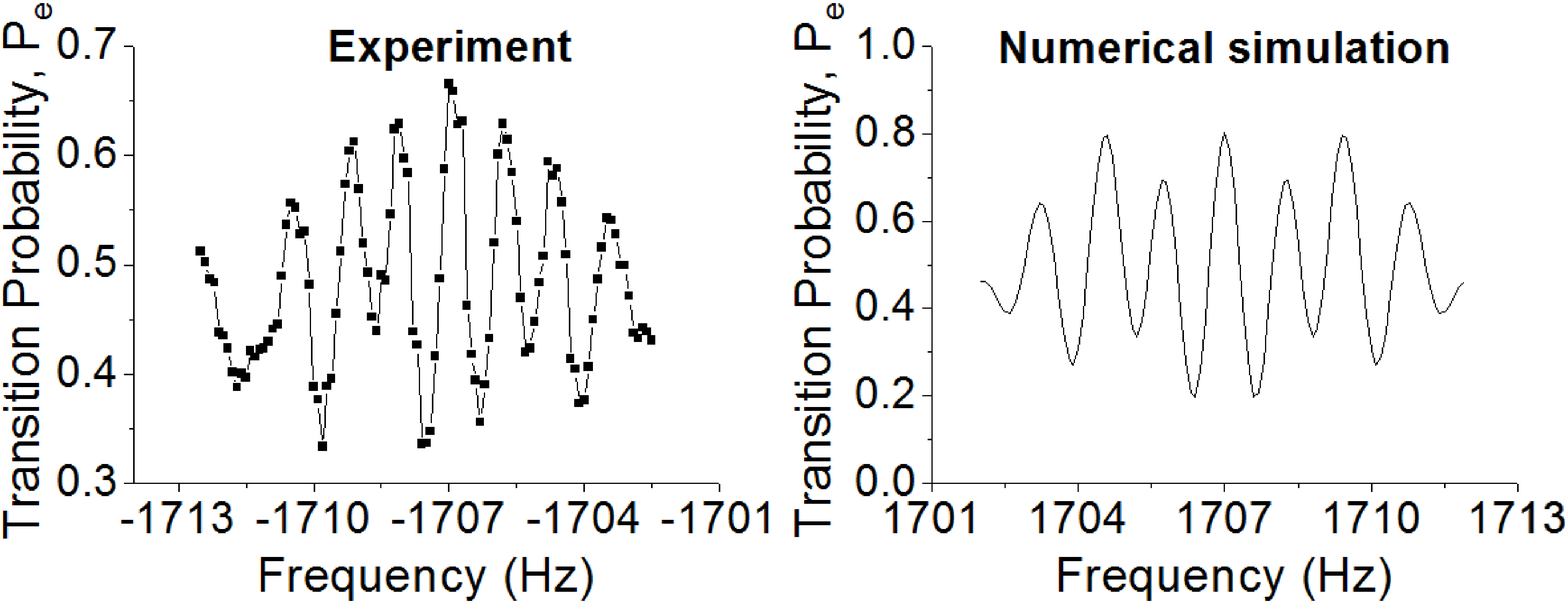}
    \caption{Fringe pattern of the accordion interferometer, for Raman pulses of ${\tau_R = 140}$~ms and separation time of ${T = 65}$~ms. Left: measured. Right: calculated. Resonant frequency inhomogeneities of about ${\sigma_{RDLS} = 2.5}$~Hz centered at ${\nu_{RDLS} = 0}$~Hz.}
    \label{Fig12_AccFringes}
    \end{center}
\end{figure}

We obtain interference fringes contrast on the ${\Delta m = -3}$ of about $30\%$ for a Raman pulse duration of ${\tau_R = 140}$~ms and a waiting time between Raman and microwave pulses of ${T = 65}$~ms (see Fig.~\ref{Fig12_AccFringes}, on the left). As in the previous case, the contrast is limited here mostly by the imperfections of the Raman transition. In addition to a Rabi pedestal, the envelope of the fringe pattern displays a modulation at twice the main periodicity. This arises from three-wave interferences, as imperfect Raman transitions also leave partial wavepackets in the initial well $m$ which finally interfere with the wavepackets placed in ${m - \Delta m}$ and ${m + \Delta m}$ with a phase difference twice smaller.

This behaviour is confirmed by a numerical simulation of the interferogram, displayed on the right in Fig.~\ref{Fig12_AccFringes}, which reproduces qualitatively the shape of the fringe pattern, but overestimates the contrast. In this simulation, the imperfection of the Raman pulses is only due to Raman differential light shift inhomogeneities, modeled with a Gaussian distribution of ${\sigma_{RDLS} = 2.5}$~Hz. For a better match of the contrast, one would need to take into account all other sources of inhomogeneities and decoherence (inhomogeneities of coupling for Raman and microwave transition, inhomogeneities of trap light shifts due to the transverse and longitudinal intensity distribution, imperfect overlap between the lasers, spontaneous emission, etc.).  

\begin{figure}[ht]
    \begin{center}
        \includegraphics[width=8.5 cm]{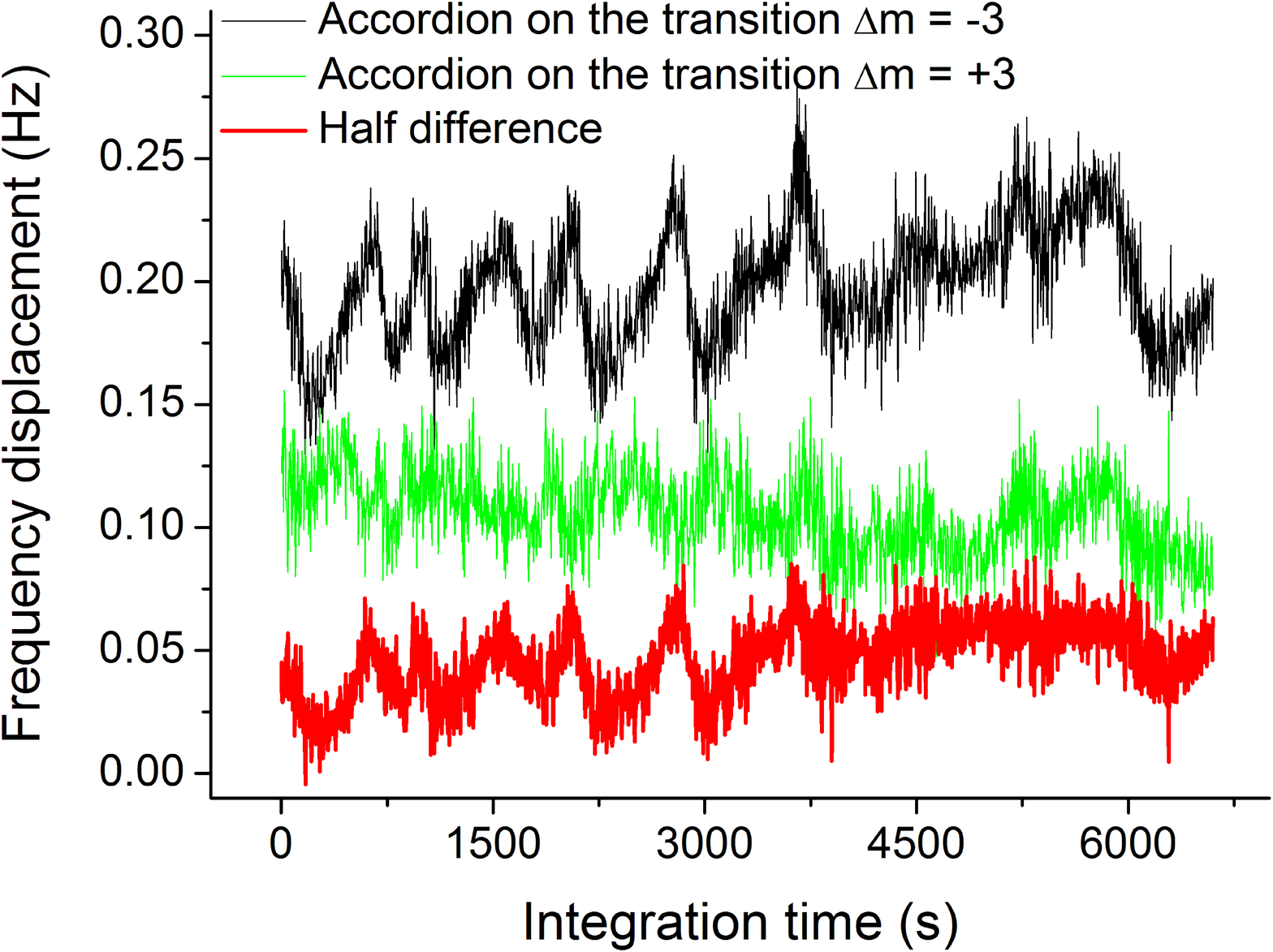}
    \caption{(Color online) Typical accordion interferometer frequency fluctuations of the central fringe measured with the Raman frequency difference $\nu_R$ set on the ${\Delta m = -3}$ (black line) and ${\Delta m = +3}$ (green line) transitions, for a Raman $\pi$-pulse of ${\tau_{R} = 80}$~ms and a time separation ${T = 205}$~ms. Half-difference (red thick line) between the two signals is presented.}
    \label{Fig13_AccTemp}
    \end{center}
\end{figure}

Using the computer controlled frequency lock described previously, we measure the frequency fluctuations of the accordion interferometer. A typical measurement is displayed in Fig.~\ref{Fig13_AccTemp}, obtained with a Raman $\pi$-pulse of $80$~ms duration tuned on the ${\Delta m = \pm 3}$ transition, and a time separation of ${T = 205}$~ms. We observe peak-to-peak fluctuations of about $40$~mHz, which correspond to a reduction by an order of magnitude compared to the Raman-Ramsey interferometer without half-difference computing. These fluctuations appear both on the ${\Delta m = -3}$ (black line) and ${\Delta m = +3}$ (green line), but with different non-stationary amplitudes inducing similar fluctuations on the half-difference (red thick line) that are not cancelled. It is important to notice that while the accordion interferometer couples the two different states $\left|m \pm \Delta m\right\rangle$ instantaneously, it is still possible to select which state will be coupled from the $\left|g, m\right\rangle$ state, either $\left|e, m + \Delta m\right\rangle$ or $\left|e, m - \Delta m\right\rangle$, by an adequate setting of the Raman frequency difference $\nu_R$ .

\begin{figure}[ht]
    \begin{center}
        \includegraphics[width=8.5 cm]{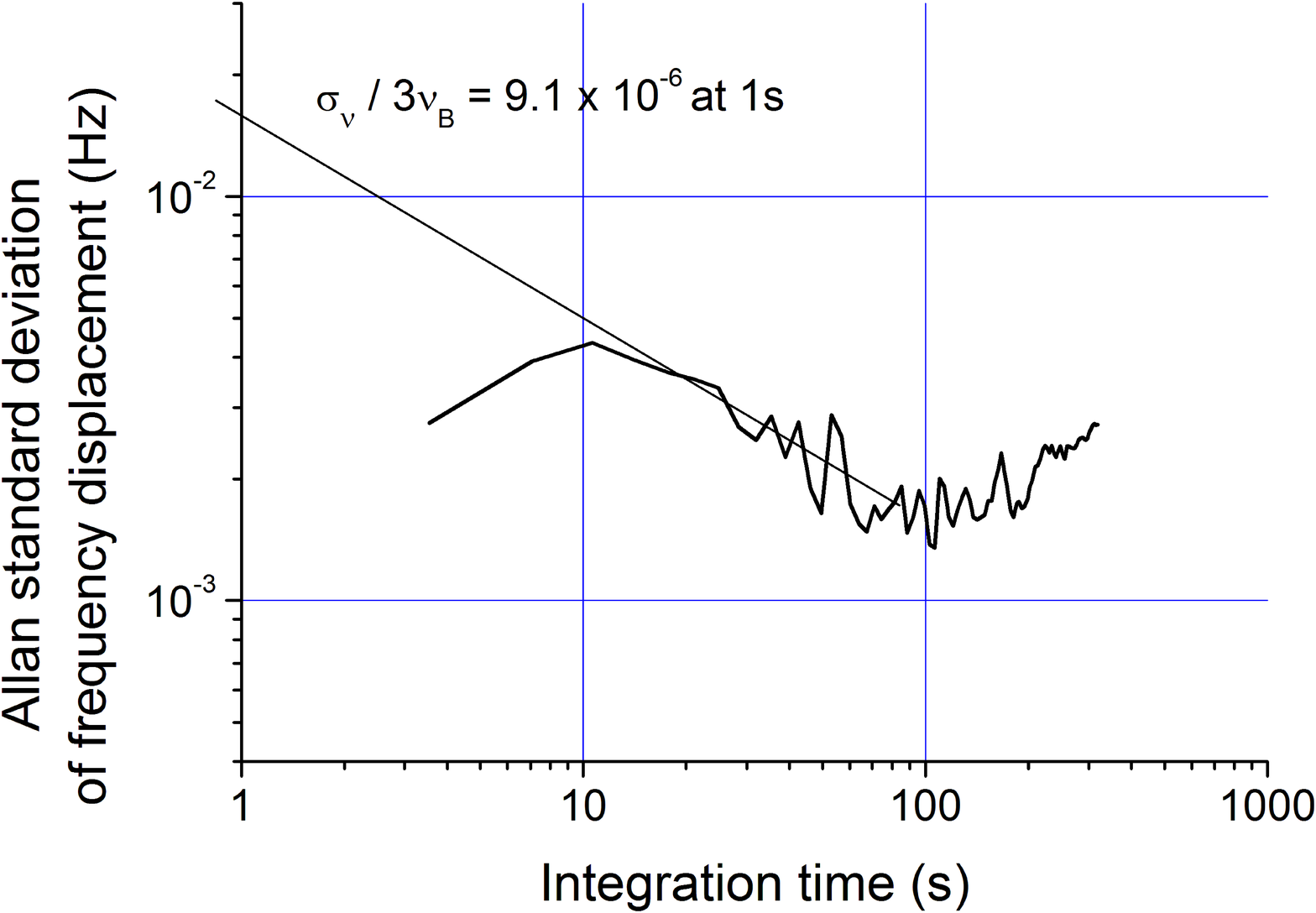}
    \caption{Best Allan standard deviation obtained from the half-difference frequency fluctuations of an accordion interferometer on the intersite transition $3 \times \nu_B$ for a Raman $\pi$-pulse of ${\tau_{R} = 140}$~ms and a time separation ${T = 175}$~ms.}
    \label{Fig14_AccAllanStd}
    \end{center}
\end{figure}

Fig.~\ref{Fig14_AccAllanStd} displays the best Allan standard deviation of the frequency fluctuations of the half-difference of our best measurement on the ${\Delta m = \pm 3}$ transition. This was obtained with the following set of experimental parameters: ${\tau_{R} = 140}$~ms and ${T = 175}$~ms. For this measurement, the residual fluctuations are of about $10$~mHz.
The short-term sensitivity on the measured frequency is ${\sigma_{\nu} = 1.55 \times 10^{-2}}$~Hz which gives a relative uncertainty on the Bloch frequency of ${\delta \nu / \nu = 9.1 \times 10^{-6}}$ at $1$~s. The gain in sensitivity is due to the twice larger spatial separation, which leads to a twice better frequency resolution.

\begin{figure}[ht]
    \begin{center}
        \includegraphics[width=8.5 cm]{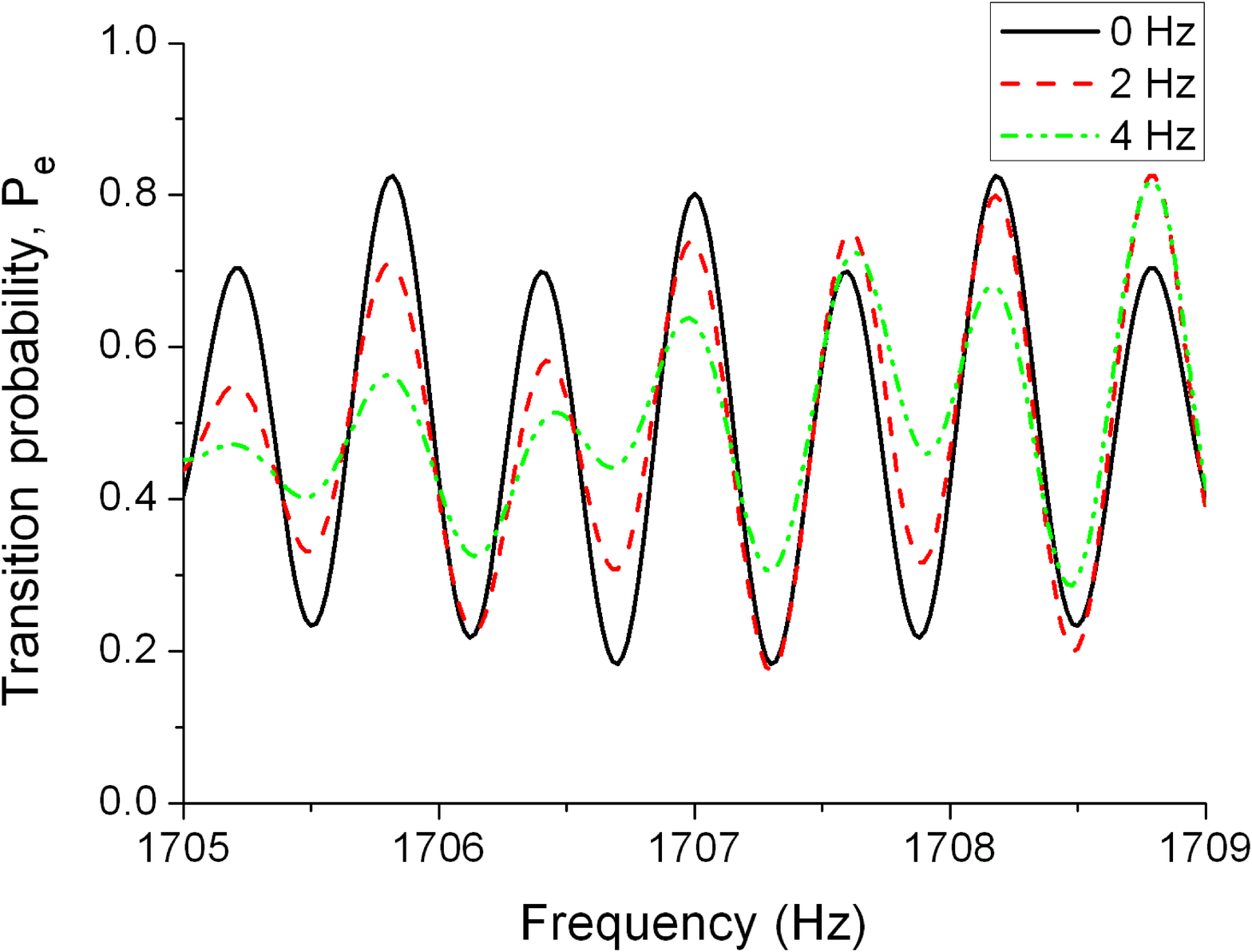}
    \caption{(Color online) Distortion of the central fringe of the ${\Delta m = + 3}$ intersite transition. The different interferometer fringe patterns are numerically calculated for different value of the mean RDLS $\nu_{RDLS}$, and for a ${\tau_R = 140}$~ms, ${T = 175}$~ms, ${\sigma_{RDLS} = 2.5}$~Hz. The central fringe of the interferometer pattern is situated at $-1707$~Hz.}
    \label{Fig15_AccLSAsymmetry}
    \end{center}
\end{figure}

Part of these fluctuations can be explained by a residual sensitivity to the Raman differential light shift. Indeed, this effect displaces the top of the Rabi envelope with respect to the central fringe. This causes a distortion of the fringe pattern and an asymmetry of the fringes, which leads to an apparent shift of the central fringe in our numerical integrator. A numerical model of the interferometer was used to study this effect. Fig.~\ref{Fig15_AccLSAsymmetry} displays the fringe pattern obtained for different values of $\nu_{RDLS}$, for the interferometer parameters ${\tau_R = 140}$~ms and ${T = 175}$~ms. We also take into account inhomogeneities of the Raman differential light shift of ${\sigma_{RDLS} = 2.5}$~Hz in the simulation, while the corresponding Rabi frequency is $3.6$~Hz. When applying our frequency measurement protocol to simulated data, we find an apparent displacement of the central fringe of $-7.6$~mHz/Hz of mean Raman differential light shift. Combined with the typical peak-to-peak fluctuations of $\nu_{RDLS}$ measured to be on the order of $50\%$ of the Rabi frequency, we expect peak-to-peak fluctuations of the position of the central fringe of $13.7$~mHz, in reasonable agreement with our best measurement described above.

The shift induced by this distortion is independent on the $\Delta m$ Raman transition, defined by the frequency difference between the two Raman lasers $\nu_R$. It thus should be well correlated in the two Bloch frequency measurements on the different ${\pm \Delta m}$ transitions realized one after each other, and should vanish when calculating the half-difference between consecutive ${\pm \Delta m}$ measurements. Nevertheless, in practice these fluctuations are still present on the half-difference with amplitude fluctuations on the order of $10$ to $40$~mHz peak-to-peak as shown on Fig.~\ref{Fig13_AccTemp}.

As for the mean value of the Bloch frequency, we find it independent on the kind of interferometer that we drive (Ramsey-Raman or accordion) at the $10^{-5}$ level in relative units.


\section{Conclusion}

We have studied two atom interferometer schemes with atoms trapped in a 1D vertical lattice: a Ramsey-type interferometer and the accordion interferometer, which constitutes a symmetric version of the former. In these interferometers, Raman transitions are used to split and recombine partial wavepackets between distinct Wannier-Stark states. This allows for a measurement of the Bloch frequency, and hence of gravity, with a relative sensitivity of ${9.1 \times 10^{-6}}$ at $1$~s and a frequency sensitivity of $15.5$~mHz at $1$~s at best. A peculiar point on this measurement is that we have reached detection noise limited performances in terms of sensitivity, with a fairly low number of atoms $<10^{5}$, which could thus be improved by increasing the number of atoms trapped in the lattice.
This level of performance is comparable to the sensitivity obtained in~\cite{PRL_Tino_SrLattice4}, where transport and delocalization of Sr atoms in a vertical lattice is induced by resonant modulation of the lattice depth. In this reference, a relative sensitivity on the Bloch frequency of ${1.5 \times 10^{-7}}$ was obtained within one hour, which corresponds to a short-term sensitivity of ${9 \times 10^{-6}}$ at $1$~s. Better performances have been obtained combining a Ramsey-Bord\'e interferometer and Bloch oscillations~\cite{PRA_Bresson}, resulting in a small interrogation distance although atoms are not trapped during the whole interferometer sequence. A relative sensitivity on the measurement of $g$ of ${2.0 \times 10^{-7}}$ was obtained after $300$~s of measurement time, which corresponds to a short-term sensitivity of ${3.4 \times 10^{-6}}$ at $1$~s. In this last experiment, the maximum separation between the wave packet amounts to $72$~$\mu$m, to be compared with our wells separation of $1.6$~$\mu$m only.
Although not competitive with the performances of free falling interferometers (${{\delta g/g \sim 10^{-8}}}$ at $1$~s), this trapped interferometer would allow for a sensitive measurement of short range forces if it is performed close to a surface. At a distance of $5~\mu$m of a (perfectly) reflecting surface, the current sensitivity would allow for reaching a statistical uncertainty in the measurement of the Casimir-Polder potential of about $1\%$ for a measurement time of only $30$~s.\\

\begin{acknowledgments}
This research is carried on within the iSense project, which acknowledges the financial support of the Future and Emerging Technologies (FET) program within the Seventh Framework Program for Research of the European Commission, under FET-Open grant number: 250072. We also gratefully acknowledge support by Ville de Paris (Emergence(s) program) and IFRAF. G.T. thanks the Intercan network and the UFA-DFH for financial support.

Helpful discussions with A. Landragin and the Inertial Sensors team, P. Wolf, S. P\'elisson, R. Messina and M.-C. Angonin are gratefully acknowledged.

\end{acknowledgments}


\end{document}